# Macroscopic coherence between quantum condensates formed at different times.

**Alex Hayat[1], Christoph Lange[1], Lee A. Rozema[1], Rockson Chang[1], Shreyas Potnis[1], Henry M. van Driel[1], Aephraim M. Steinberg[1], Mark Steger[2], David W. Snoke[2], Loren N. Pfeiffer[3] and Kenneth W. West[3]**

[1]*Department of Physics, Centre for Quantum Information and Quantum Control, and Institute for Optical Sciences, University of Toronto, Toronto, Ontario M5S 1A7, Canada*

[2]*Department of Physics and Astronomy, University of Pittsburgh, Pittsburgh, Pennsylvania 15260, USA*

[3]*Department of Electrical Engineering, Princeton University, Princeton, New Jersey 08544, USA*

**<u>Abstract</u>**

We demonstrate macroscopic coherence between quantum condensates generated at different times, separated by more than the particle dephasing time. This is possible due to the dressed light-matter nature of exciton-polaritons, which can be injected resonantly by optical excitation at well-defined momenta. We show that the build-up of coherence between condensates depends on the interaction between the particles, particle density, as well as temperature despite the non-equilibrium nature of the condensate, whereas the mass of the particles plays no role in the condensation of resonantly injected polaritons. This experiment also makes it possible for us to measure directly the large nonlinear phase shift resulting from the polariton-polariton interaction energy. Our results provide direct evidence for coherence between different condensates and demonstrate a new approach for probing their ultrafast dynamics, opening new directions in the study of matter coherence as well as in practical applications such as quantum information and ultrafast logic.

**<u>Introduction</u>**

Interacting bosons can form coherent states of matter by generating quantum condensates[1,2,3,4,5]. Spatial interferometry has been crucial for the demonstration of macroscopic coherence in cold-atom condensates[6, 7] and in the rapidly-developing field of exciton-polariton condensates[8,9,10]. Such measurements have been performed on steady-state systems, and therefore unable to access the non-equilibrium dynamics of the macroscopic coherence. Macroscopic coherence is one of the key properties of quantum condensates[1-5] and superfluids[11], and has been widely studied in various steady-state systems[6-9]. Exciton-polaritons, which emerge from strong light-matter coupling in semiconductor microcavities[12], exhibit a combination of extremely low effective mass and strong polariton-polariton interactions, leading to a wide range of physical phenomena including high-temperature condensation[3-5] and superfluidity[13,14,15,16,17]. A unique advantage of exciton-polariton condensates is that they can be directly accessed with light, so that various properties such as spin, energy, momentum and phase of the excitation photons are mapped onto polaritons. This attribute has been employed in resonant imprinting of coherence on polariton condensates to generate quantized vortices[16] and to demonstrate long temporal coherence in steady-state condensates using narrowband continuous-wave lasers[18].

We demonstrate macroscopic coherence between two dynamic polariton condensates generated at times separated by more than the polariton dephasing time, $T_2$. The polariton populations are injected by ultrafast laser pulses at different transverse momenta, with the bandwidth determined by $T_2$ – slightly narrower than the excitation pulse bandwidth. Stimulated polariton-polariton scattering has been shown to be the main mechanism responsible for generation of condensates in quasi-equilibrium polariton systems[19]. In our pulsed resonantly-injected condensates, stimulated polariton-polariton scattering results in spectral narrowing of the dynamic condensates and thus in an increased decay time of the macroscopic coherence, $T_{2c}$. We study the dynamics of the macroscopic coherence by monitoring the interference fringe visibility between two condensates injected at different times. We show that for exciton-like polaritons stronger interaction results in larger condensate fraction and thus in longer decay time of the macroscopic coherence – much longer than $T_2$. Photon-like polaritons, on the other hand, result in lower stimulated scattering rates and thus in generation of a smaller condensate fraction, as well as in a shorter polariton lifetime, $T_1$.

**Results**

In resonant polariton injection, a finite energy range is populated by polaritons with the initial bandwidth determined by the laser and the polariton linewidth. The decay time of macroscopic coherence in our experiments due to condensation, $T_{2c}$, is, however, much longer than the one defined by the injected bandwidth. Without polariton condensation the coherence time is limited by the polariton dephasing time, $T_2$. We have measured this dephasing time in our samples directly to be $T_2 \sim 1$ ps using ultrafast pump probe spectroscopy[20], by observing the decay of the coherent oscillations[21] (Fig. 1a). The corresponding dephasing-limited linewidth of ~0.6 meV, also observed in reflectivity measurements (supplementary Fig. S1), is the energy bandwidth of the injected polariton population in our experiment. It is slightly narrower than the excitation pulse bandwidth. Therefore, without condensation, the interference visibility must decay rapidly within $T_2 \sim 1$ps. When the density of injected carriers is high enough for condensation to occur, the bandwidth narrows and the coherence time of the condensate $T_{2c}$ exceeds the initial coherence time of the non-condensed polaritons, $T_2$. Thus in the condensed case macroscopic coherence can be observed between condensates formed at times separated by much longer than $T_2$.

Our experiments were performed in a He-flow microscopy cryostat, on a strongly-coupled AlGaAs microcavity[20] (details in Supplementary Methods). The output of the ~775nm, 81 MHz repetition rate Ti:sapphire mode-locked laser was spectrally filtered with a narrow band pass filter (BPF) to result in ~700 fs pulses – slightly broader than the polariton linewidth. The pulse train was split into two paths with a controllable relative time delay, $\Delta t$, and recombined on the sample at different angles (Fig. 1 b, 1c). The central wavelength of the pulses was tuned by tilting the BPF, and the incidence angles determined the injected polariton in-plane momenta, $k_1$ and $k_2$. Each pulse from the pair injected a polariton population centred at an in-plane momentum given by the excitation pulse angle, and the interference between the two generated condensates formed an interference fringe pattern. The luminescence from the polariton decay in the condensate was imaged onto a charge coupled device (CCD) camera to record the condensate interference pattern (See Supplementary Methods).

The first set of interference measurements was performed at a temperature of 10 K with high average excitation intensity (~10 mWcm$^{-2}$), and close to optimal cavity-exciton coupling, where both lower polariton (LP) and upper polariton (UP) have nearly equal exciton fractions. The LP was excited resonantly at time $t = 0$, forming a dynamic condensate at an in-plane momentum $k_1 \approx$ -0.7 $\mu$m$^{-1}$. At a time $t = \Delta t$, much longer than polariton T$_2$, another population was injected resonantly at an in-plane momentum $k_1 \approx$ +0.7 $\mu$m$^{-1}$. The two formed condensates have substantial spatial overlap, resulting in an interference pattern (Fig. 2). This interference is a signature of macroscopic coherence between the two condensates. Spatial interference in polariton condensates has been employed recently to demonstrate coherence[8-9], however the observation presented here introduces two conceptually new aspects. The macroscopic coherence in all previous experiments was demonstrated on a single condensate with spatially dependent excitation[8-9]. So far, no experiment was performed on two or more distinct condensates.

Contrarily, our experiments study the interference of two distinct condensates, where initially only one condensate exists, while the second one is formed at a later time. Moreover, we show the dynamics of condensate formation using pulsed excitation in the cross-coherence between two condensates generated at times separated by longer than $T_2$.

The LP linewidth at optimal coupling in our sample is about 0.6 meV, corresponding to $T_2 \sim 1$ ps. For LP condensates, the coherence persists for longer than $T_{2c} \sim 10$ ps (Fig. 2 LP). We performed a similar experiment on two condensates formed in the UP branch. The UP condensate coherence time was measured to be $T_{2c} \sim 3$ ps, which is longer than $T_2 \sim 1$ ps, but shorter than $T_{2c}$ of the LP condensate (Fig. 2 UP).

The long decay time of macroscopic coherence $T_{2c}$ observed here originates from an emitter with less than a tenth of the initially injected bandwidth (Fig. 2). This spectral narrowing results from dynamic condensation of polaritons and becomes more significant at higher excitation intensities. The shorter coherence time in the UP condensate relative to that of the LP is attributed to the smaller condensate fraction, resulting from the faster population decay. Condensate formation results from bosonic stimulated scattering, and in polariton condensation, polariton-polariton scattering has been shown to be the dominant mechanism[22]. The rate of polariton-polariton scattering depends strongly on the excitonic fraction, $\left| X_{k_\parallel} \right|^2$, of the polaritons. Therefore a smaller excitonic Hopfield coefficient, $X_{k_\parallel}$, leads to a smaller condensate fraction. Smaller $X_{k_\parallel}$ also results in shorter lifetime of the polaritons, $T_1$, which also affects the visibility decay in our pulsed experiments (see Supplementary Methods).

We measured the decay of condensate coherence for several different excitonic fractions of the LP. This was achieved experimentally by controlling the cavity-exciton detuning, $\Delta E$, which changes continuously across the sample (see Methods). Close to the optimal cavity-exciton detuning, $\Delta E \approx 0$, the macroscopic coherence between the two condensates persists for longer than 10 ps indicating a large condensate fraction (Fig. 3 a), and the visibility decay is limited only by the condensate lifetime, $T_1$ – much longer than polariton dephasing, $T_2$. For a slightly red-detuned cavity, $\left(\Delta E \approx -6 \quad meV\right)$, the LP is more photon-like, with a smaller excitonic coefficient $X_{k_\parallel}$ resulting in shorter condensate lifetime, $T_1$, and lower polariton-polariton scattering rate. The lower scattering rate leads to a smaller condensate fraction, which manifests itself in shorter coherence times. An even larger red-detuning of the cavity from the exciton $\left(\Delta E \approx -17 \quad meV\right)$ yields an even smaller condensate fraction and shorter coherence times, approaching the limit of a photon-like polaritons with no condensation. However, for large excitonic fraction, the fact that the measured visibility decay time is significantly longer than $T_2$, is clear evidence of condensation and line narrowing. In addition to decreasing the polariton scattering rates and $T_1$, the smaller excitonic fraction results in a smaller LP effective mass $m_{LP} = \left(\left|X_{k_\parallel}\right|^2 \Big/ m_X + \left(1 - \left|X_{k_\parallel}\right|^2\right) \Big/ m_C\right)^{-1}$, where $m_X$ and $m_C$ are the exciton and the cavity-photon effective masses, respectively. In equilibrium (or quasi-equilibrium) condensates, smaller mass leads to higher condensate fractions, since the coherence length of the particle wavefunctions (thermal wavelength) must be longer than roughly the inter-particle spacing for condensation to occur[23]. In highly non-equilibrium condensates, formed from resonantly injected polaritons, such as the ones in our experiments, however, the effective mass does not affect the polariton coherence length. The LP coherence is determined directly by the injected momentum bandwidth, and condensation onset depends mainly on the injected polariton density and on the polariton-polariton scattering rate. We modelled the injected non-equilibrium condensate dynamics by the generalized Gross-Pitaevskii equation[24] coupled to the reservoir, with simulation parameters scaled within an order of magnitude of the typical values (see Methods), and the calculated dependence of the condensate coherence agrees well with the measurements (Fig. 3 a).

In order to demonstrate the dependence of the condensate fraction on the polariton density, we studied the dependence of the macroscopic coherence on the excitation intensity. Holding $X_{k_{\parallel}}$ fixed, increasing the pump intensity should result in a larger condensate fraction, and thus higher macroscopic coherence at a given delay time of $\Delta t = 6\,ps$ between two generated condensates. Our intensity-dependent condensate interference visibility verifies this property (Fig. 3 b). The dynamic condensate is far from thermal equilibrium, and its temperature is not well-defined. Nevertheless, the temperature of the semiconductor lattice can be controlled, and high lattice temperature has been shown previously to cause condensation deterioration in quasi-equilibrium systems[25]. Our experimental results show that at higher lattice temperature, higher pump intensities are required to reach a given coherence level (Fig. 3 b) due to line broadening by phonons. Increasing the lattice temperature from 10 K to 50 K requires a ten-fold increase in pump intensity for condensation onset, comparable to recently reported results for incoherently generated polariton condensates[26].

We used the dynamics of macroscopic coherence accessible in our experiments to demonstrate the effect of the polariton interactions on the interference between two condensates. After the excitation pulse, the energy of the condensate $\Psi$ is blue-shifted due to the mean-field interactions by $g|\Psi|^2$ (see Methods), an energy shift which increases with increasing pump intensity, $I$, demonstrated previously in spectral measurements[3,8,9]. In our time-dependent interference experiments, the blue shift of the condensate formed first, will result in its phase shifted by the generation time of the second condensate, the phase shift $\Delta\varphi$ increasing with the blue shift $g|\Psi|^2(t,I)$ and the delay time $\Delta t$:

$$\Delta\varphi = \int_{0}^{\Delta t} g|\Psi|^2(t,I)dt. \qquad (1)$$

Therefore for constant delay time, $\Delta t$, increasing the pump intensity $I$ results in a larger phase difference between the two interfering condensates, visible as a spatial shift in the interference pattern.

We observed this nonlinear phase shift in our interference experiments (Fig. 4). At a delay time of $\Delta t = 6\,ps$, increasing $I$ while keeping all other parameters constant results in a shift of the interference fringes with nearly linear dependence on $I$ (Fig. 4 a). At a longer $\Delta t = 16\,ps$, a larger shift of the interference pattern is observed due to a larger phase accumulation $\Delta\varphi$ (Fig. 4 b). The phase shift dependence is in good agreement with our calculations (Fig. 4 c). From a nonlinear optics perspective, this phase shift represents an extremely strong nonlinear process. The observed phase shift here is on the scale of $10^{-4}$ rad/photon – two orders of magnitude larger than the recently reported results on cold-atom nonlinearities using electromagnetically-induced transparency[27]. Such nonlinearities are at the forefront of atomic physics research aimed at quantum nondemolition measurements and quantum computing[28]. In cold-atom systems, intensity-dependent phase shifts can be obtained for nearly-resonant transitions, but the maximum shift is limited by the atom density. In polariton condensates, the large nonlinearity stems from the inherent light-matter dressed states in the system. This optical nonlinearity is fundamentally different because it is based on the strong Coulomb interaction between charged particles in the dressed states, in contrast to high-order multi-photon transitions in conventional nonlinear optics.

Our results shed new light on the dynamics of macroscopic coherence in matter, which is demonstrated here between two separate condensates formed at different times, allowing the study of the unique ultrafast behaviour of nonequilibrium condensates, and the development of new devices for future technologies.

## Methods:

**Theoretical modelling:**

In the theoretical model, the pump pulse excites a population of polaritons – the reservoir, $n_R$, over an energy bandwidth that corresponds to the injected optical bandwidth given by $T_2$. Following the injection, polaritons scatter from the reservoir into the condensate $|\Psi|^2$. The calculation of condensate coherence dynamics (Fig.3), as well as the nonlinear phase shift (Fig. 4), were performed using the generalized Gross-Pitaevskii nonlinear equation[24]

$$i\frac{\partial \Psi}{\partial t} = \left[ -\frac{\hbar \nabla^2}{2m_{LP}} + \frac{i\left[ R(n_R) - \gamma \right]}{2} + g|\Psi|^2 + g_R n_R \right]\Psi \qquad (2)$$

where $\gamma = 1/T_1$ is the polariton loss rate. The polariton-polariton interaction within the condensate is modelled by a mean-field coupling constant, $g$, while the interaction with the polariton reservoir density given by a coupling constant $g_R$. The dynamics of the reservoir is described by

$$\partial n_R / \partial t = -\gamma n_R - R\left(n_R\right)\left|\Psi\right|^2 - D\nabla^2 n_R \qquad (3)$$

where $D$ is the reservoir polariton diffusion constant and $R\left(n_R\right)$ is the scattering rate of polaritons from the reservoir into the condensate. The coupled equations (2) and (3) account for stimulated scattering between the reservoir and condensate neglecting temperature effects.

**Control of polariton interaction strength:**

For initial and final polariton momenta, $k_i$, $k_j$ and $k_l$, $k_m$, the polariton-polariton scattering matrix element is given by[19]:

$$M \propto \sum_{k_i, k_m} \frac{X_{k_i} X_{k_j} X_{k_l} X_{k_m}}{\left|k_i - k_m\right|}. \qquad (4)$$

The detuning $\Delta E\left(k_\parallel\right) = E_C\left(k_\parallel\right) - E_X\left(k_\parallel\right)$ of the cavity energy $E_C\left(k_\parallel\right)$ from the exciton energy $E_X\left(k_\parallel\right)$, can be controlled by changing the excitation location due to the tapered layer thickness across the sample. This detuning determines the polariton Hopfield coefficient for the exciton fraction $\left|X_{k_\parallel}\right|^2 = \frac{1}{2}\left(1 + \Delta E\left(k_\parallel\right) \Big/ \sqrt{\Delta E\left(k_\parallel\right)^2 + \left(\hbar\Omega_0\right)^2}\right)$ of the lower polariton (LP), where $\Omega_0$ is the Rabi frequency of exciton-photon coupling. Experimentally, $\Delta E\left(k_\parallel\right)$ was determined by reflection spectroscopy from the sample (Supplementary Fig. S1).

**Figure captions:**

**Figure 1. Time-dependent polariton experiments** **(a)** Measurement of polariton dephasing time, $T_2$. Differential reflectivity $\Delta R/R$ spectra in ultrafast pump-probe spectroscopy. $T_2 \sim 1$ ps is observed in the decay of the coherent oscillations. **(b)** LP and UP dispersions with a schematic representation of the resonant condensate excitation at different inplane momenta, $k_1$ and $k_2$, and at different times $t = 0$ and $t = \Delta t$. **(c)** Schematic diagram of the experimental setup.

**Figure 2. Interference pattern for different delay times. (LP)** LP resonant excitation **(UP)** UP resonant excitation

**Figure 3. Macroscopic coherence measurements. (a)** as a function of delay time for various cavity-exciton detunings. The solid lines are the calculated dependence. **(b)** as a function of excitation intensity for various lattice temperatures and for $\Delta t = 6 \, ps$. The solid lines are power-law fits as a guide to the eye. The black line is the calculated dependence with the maximum visibility as a fitting parameter.

**Figure 4. Nonlinear phase shift measurements** **(a)** Interference pattern shift with increasing pump power at $\Delta t = 6 \, ps$ delay. The dashed lines are a guide to the eye following the fringe pattern minima **(b)** Interference pattern shift with increasing pump power at $\Delta t = 16 \, ps$ delay. **(c)** Nonlinear phase shift vs. pump power at $\Delta t = 16 \, ps$ delay. The solid red line is the calculated dependence.

**Figure 1**

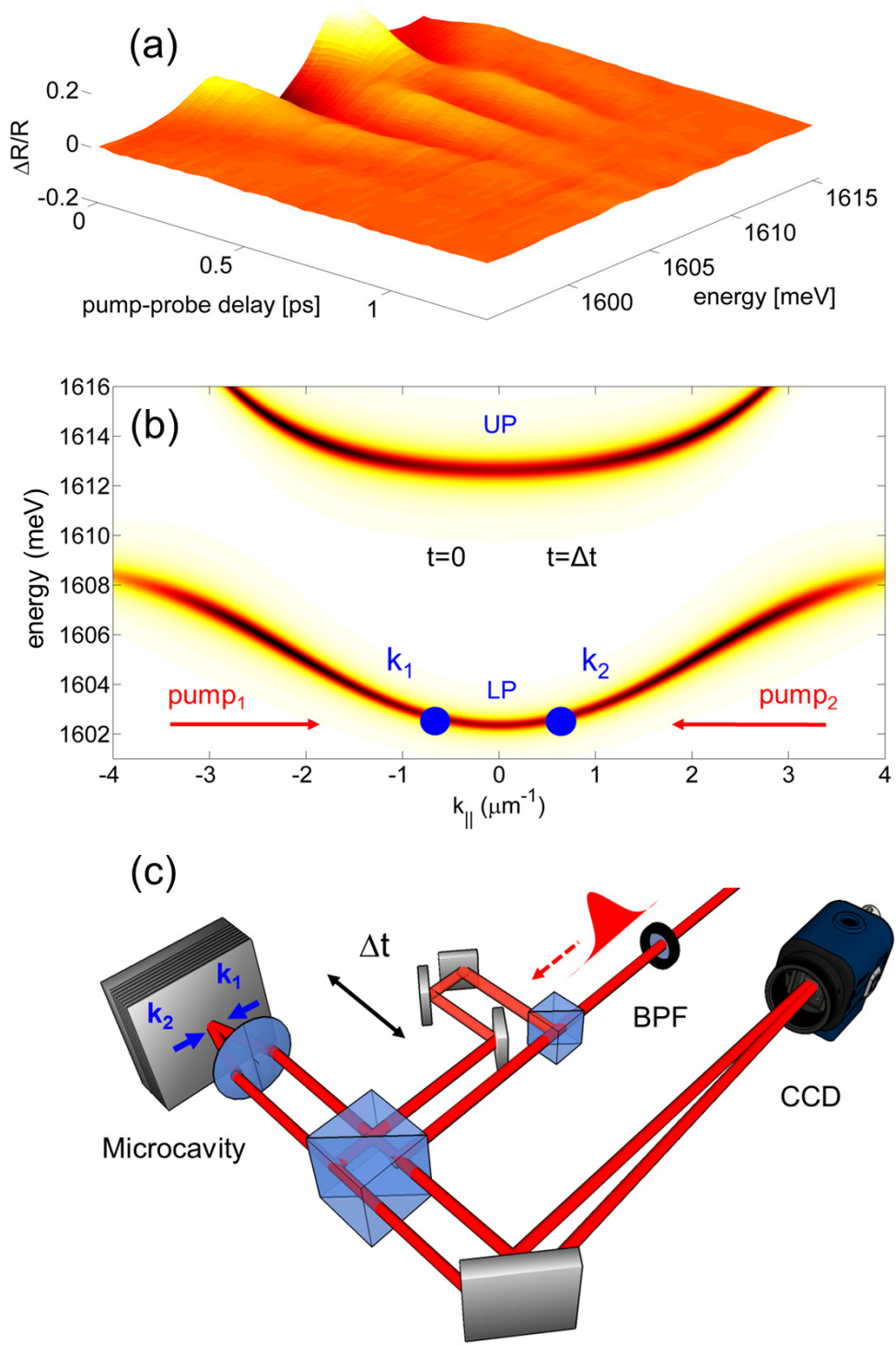

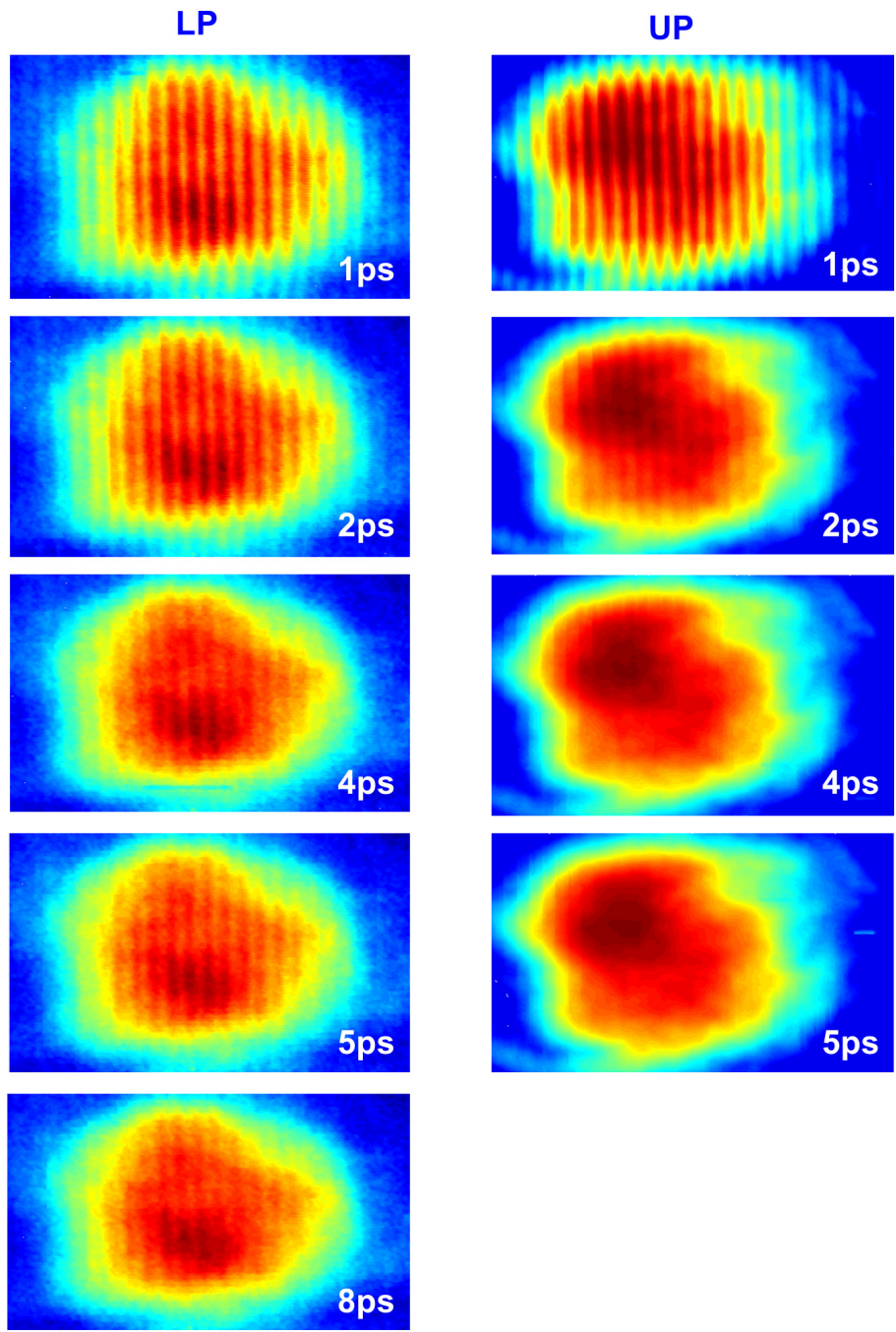

**Figure 3**

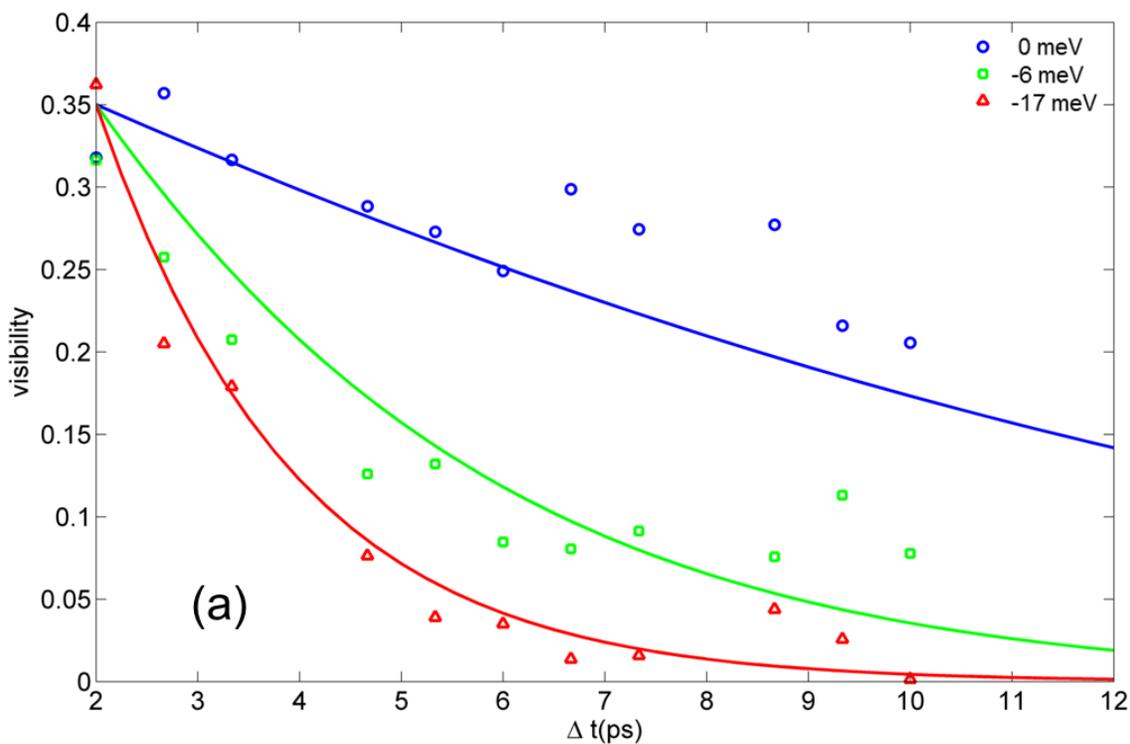

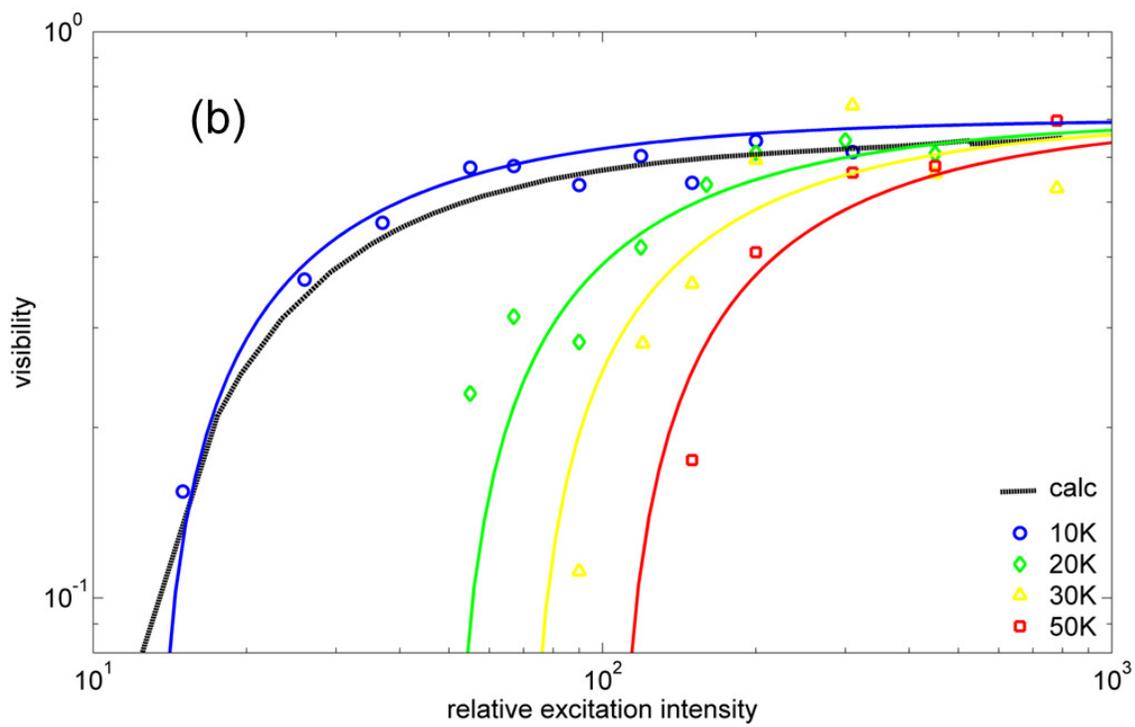

## Figure 4

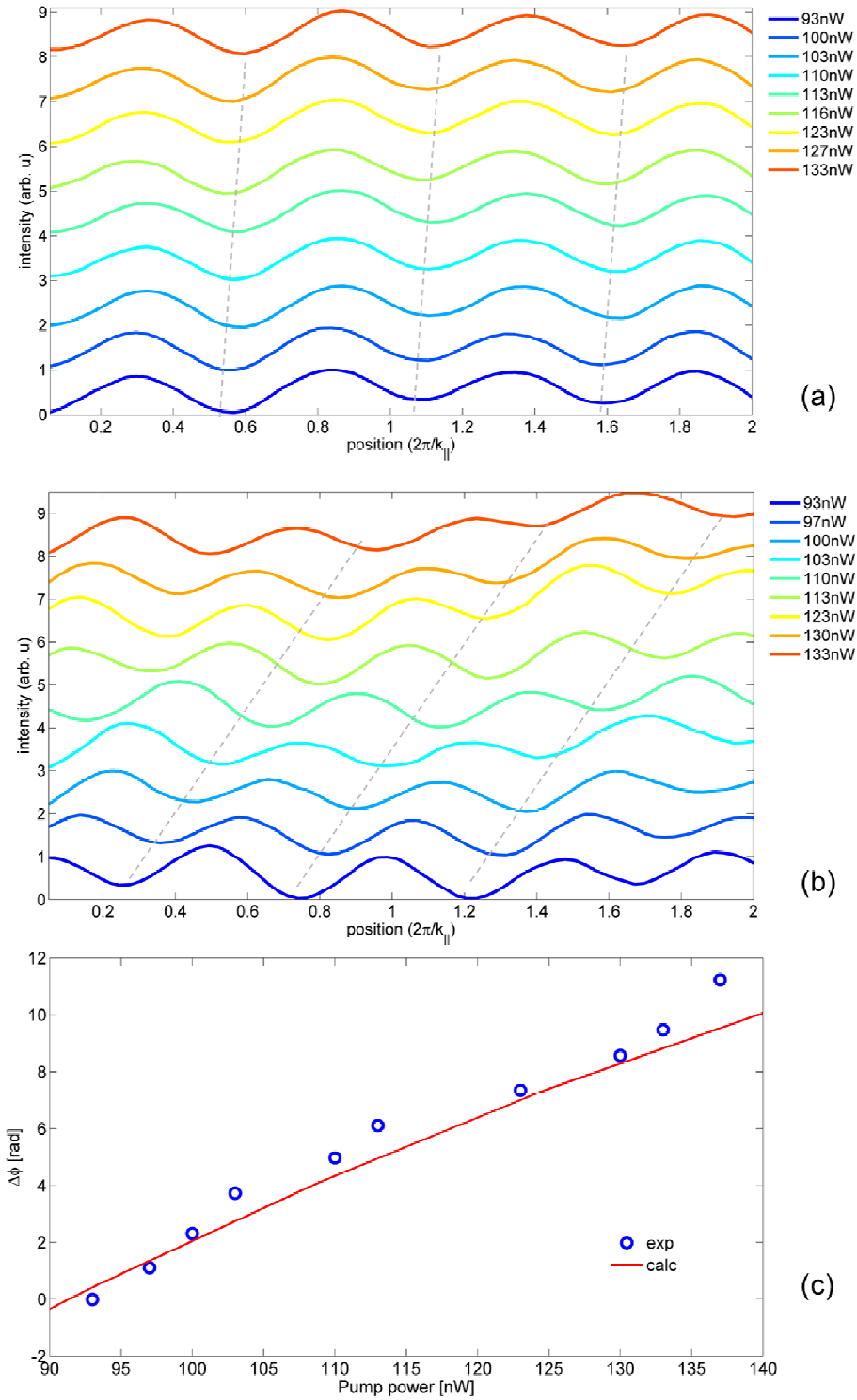

# Supplementary Information

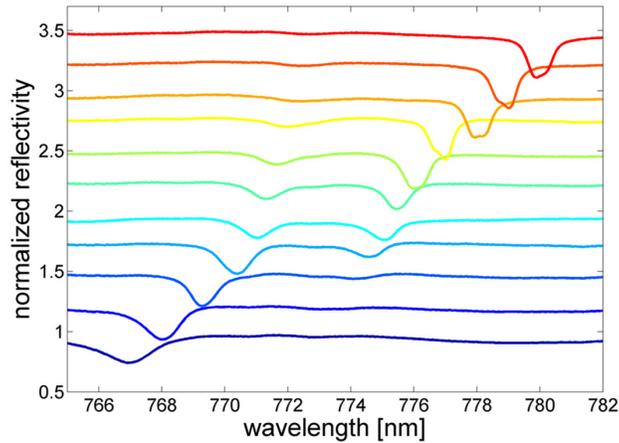

**Supplementary Figure S1 | Microcavity reflectivity measurements:** Normalized reflectivity along the taper direction, vertically offset for clarity. The typical anti-crossing is observed at the cavity-exciton resonance of about 773 nm.

**Supplementary Methods**

**Device Fabrication and Characterization:**

The strongly coupled $3\lambda/2$ microcavity, grown by molecular beam epitaxy, consists of a $\lambda/2$ AlAs layer between a bottom 20-layer-pair $Ga_{0.8}Al_{0.2}$/AlAs distributed Bragg reflector (DBR), and a top 16-pair DBR. A single 6.5 nm GaAs QW was placed at each of the 3 antinodes of the microcavity with AlAs barriers. Our structure is similar to the one used in [20]. The two AlAs barriers of the central QW are parts of the 63 nm thick $\lambda/2$ layer, and the other two QWs have a 10nm AlAs barrier and a 59 nm AlAs barrier. The thickness of the layers was tapered across the sample to allow tuning of the cavity resonance across the exciton resonance by probing at different locations on the sample.

**Visibility extraction**

The visibility measurement is averaged over the duration of both condensate lifetimes, $T_1$. Assuming that the sizes of both condensates are equal, the instantaneous intensity on the CCD is

$$I_t(t) = I(t) + I(t - \Delta t) + 2\sqrt{I(t)I(t - \Delta t)} \cos(2kx) e^{-\Delta t / T_{2c}} + noise \quad (S1)$$

where $I(t)$ is the intensity corresponding to each individual condensate alone, $T_{2c}$ is the condensate coherence time, $\Delta t$ is the delay between the two injection times, and $\cos(2kx)$ is the spatial interference fringe pattern. The normalized instantaneous visibility at time $t$ is then

$$V(t, \Delta t) = \frac{2\sqrt{I(t)I(t - \Delta t)}}{I(t) + I(t - \Delta t) + noise} e^{-\Delta t / T_{2c}}. \quad (S2)$$

The CCD records the time-integrated intensity $\int I_t(t) dt$, so that the extracted visibility is an integral of normalized instantaneous visibility weighted by the instantaneous intensity or,

$$\bar{V}(\Delta t) = \int V(t, \Delta t)\left(I(t) + I(t - \Delta t) + noise\right) dt. \quad (S3)$$

Therefore, in the limit of large $T_{2c}$, the extracted visibility is limited by the polariton lifetime $T_1$ - much longer than $T_2$, which determines the injected bandwidth as mentioned in the text. This long $T_{2c}$ limit occurs when the exciton fraction of the polaritons is large. For smaller exciton fraction, both $T_1$, and $T_{2c}$ become shorter, and both affect the visibility decay.

The integration time of the CCD (~ms) is much longer than the relevant timescales such as $T_1$, so the extracted visibility $\bar{V}(\Delta t)$ is effectively an average value (Eq. S3). The time-dependent blue-shift can reduce the extracted visibility slightly, however the effect is not significant because in the integrated $\bar{V}(\Delta t)$ measurement the lower-intensity points in time contribute less, and $\bar{V}(\Delta t)$ is determined mainly by the high-intensity parts of the emission as is seen in our calculations.

The absolute value of visibility depends on temperature because at each temperature a different spot is excited on the sample to keep the same excitonic fraction of polaritons. Therefore Fig. 3b shows normalized visibility so that at very high intensities all temperature curves saturate at the same value.